\documentclass[a4paper,11pt]{article}
\usepackage{pos}

\title{MSHT Updates 2024}

\author*[a]{R.S. Thorne}
\author[b]{T.Cridge}
\author[a]{L. Harland-Lang}

\affiliation[a]{Department of Physics and Astronomy, University College London, London, WC1E 6BT, UK }

\affiliation[b]{Deutsches Elektronen-Synchrotron DESY, Notkestr. 85, 22607 Hamburg, Germany }

\emailAdd{robert.thorne@ucl.ac.uk}
\emailAdd{thomas.cridge@desy.de}
\emailAdd{l.harland-lang@ucl.ac.uk}

\abstract{We present various recent updates related to the MSHT20 PDF sets, and particularly to the MSHT20 approximate N$^3$LO (aN$^3$LO) PDFs. These include an investigation of including either inclusive jet data or dijet data, the detailed treatment of $Z\,p_T$ data, and the impact on the gluon PDF in particular. We also very briefly summarize the determination of the strong coupling via an aN$^3$LO global PDF fit, finding $\alpha_S(M_Z^2)=0.1170\pm 0.0016$, and of including QED corrections in the aN$^3$LO analysis. Finally we consider the preliminary finding associated with the effect of improved knowledge of the N$^3$LO splitting functions on the aN$^3$LO PDFs. In particular we note a moderate impact on the gluon distribution, i.e. within the uncertainty, and a more limited impact on the gluon-gluon luminosity unless strong rapidity cuts are imposed. Hence, we conclude that while details of our aN$^3$LO PDFs will change with more precise splitting functions, only relatively minor effects on predictions are likely to result. }

\FullConference{%
  DIS2024\\
  8-12 April 2024 \\
  Grenoble, France
}

\begin{document}
\hspace*{\fill} DESY-24-115
\maketitle

\section{Updates following the MSHT20 Global Fit}

\begin{figure}
\begin{center}
\includegraphics[scale=0.64]{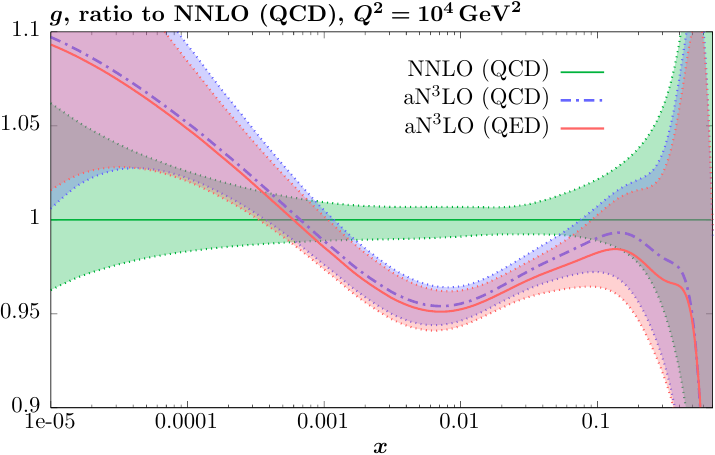}
\hspace{0.2cm}
\includegraphics[scale=0.34]{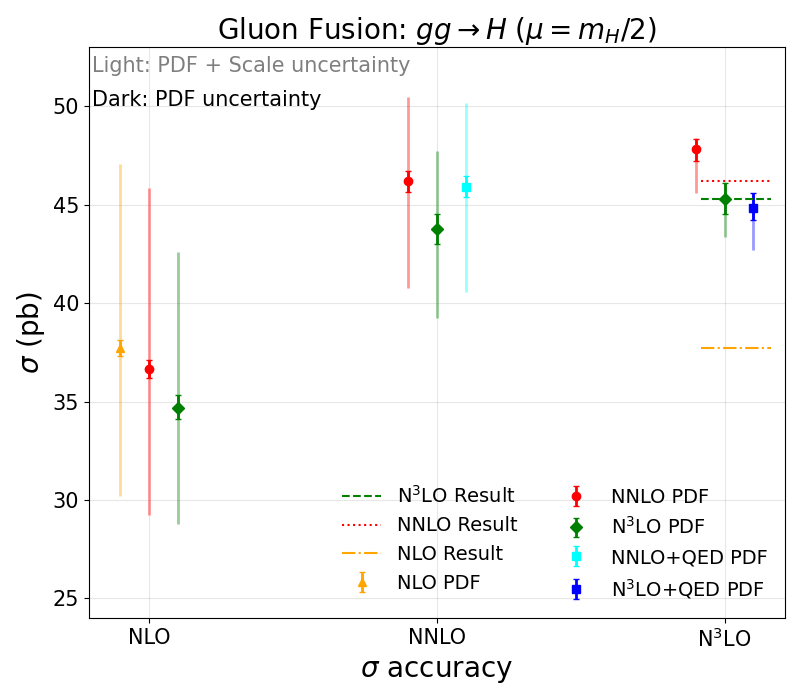}
\caption{\sf The NNLO QCD-only gluon distribution compared to the aN$^3$LO QCD-only and the aN$^3$LO QCD+QED gluon distributions as a function of $x$ at $Q^2=10^4$~GeV$^2$ (left). The prediction for the cross section for the production of a Higgs boson by gluon-gluon fusion using various combinations of cross section perturbative order and PDFs produced a various perturbative orders (right). }
\label{fig:n3logluon}
\end{center}
\end{figure}

There have been numerous updates in the MSHT series of PDF studies following the last major update of the global fit in 2020 \cite{Bailey:2020ooq}. Soon afterwards there appeared a dedicated study of the variation of the PDFs with different values of heavy quark masses and of the strong coupling constant $\alpha_S(M_Z^2)$ \cite{Cridge:2021qfd}, with a determination of the best-fit value and uncertainty of the latter giving $\alpha_S(M_Z^2)=0.1174 \pm 0.0013$ at NNLO. 
The inclusion of QED effects and a photon distribution were additionally presented in \cite{Cridge:2021pxm}. Both of these were relatively standard extensions which have been made on previous major updates of our PDF sets. However, more recently we  used the MSHT framework as a basis for producing the first approximate N$^3$LO (aN$^3$LO) PDFs \cite{McGowan:2022nag} using the large amount of information already available at this time on the N$^3$LO corrections to splitting functions, transition matrix elements and hard cross sections, and parameterising the missing information in terms of nuisance parameters which provide a theoretical uncertainty. This was a much more innovative development than more standard extensions to the global determinations of NNLO in perturbative QCD PDF sets, and some significant changes in PDFs were seen at aN$^3$LO, most notably a decrease in the gluon compared to NNLO at $x \sim 0.01$, as shown in Fig.~\ref{fig:n3logluon} (left).
A similar study has recently been performed by the NNPDF group \cite{NNPDF:2024nan}. Further work has in addition been performed on benchmarking the two evolution codes, investigating the reasons for differences in evolution and examining the effects of updates to the information available on the splitting functions \cite{Falcioni:2023luc,Falcioni:2023vqq,Moch:2023tdj} since the PDFs were first obtained \cite{Cooper-Sarkar:2024crx,Andersen:2024czj}. In general both groups 
see similar trends in the aN$^3$LO to NNLO comparison of PDFs, though somewhat smaller in the NNPDF case, but codes agree when using the same inputs and progress has been made on understanding differences between the two groups' results, due to both evolution and fit procedures, as will be discussed more later. 

Following our introduction of the aN$^3$LO PDF sets there have been a number of wider studies which include these. It is a long standing question as to whether it is more appropriate to include inclusive jet or dijet data in global fits to PDFs, since the use of both simultaneously is often not possible without double counting. Historically, availability of more inclusive jet data, and some poor fit qualities to dijet data at NLO, have led to the former being the default for PDF studies. There was a first investigation of the effects of using dijet data  instead in \cite{AbdulKhalek:2020jut}. We have recently performed a similar study \cite{Cridge:2023ozx}, and moreover we addressed  the case of the aN$^3$LO PDFs. At NNLO we agree with the general observations of \cite{AbdulKhalek:2020jut}; in particular we see a much improved fit quality at NNLO, indeed rather better than that for inclusive jet data, and also see less tension between dijet data and the other data in the global fit than we do for inclusive jet data. In addition we find that the fit quality improves for dijet data when either electroweak or N$^3$LO corrections are added, which is not so clearcut for the inclusive jet case. In the same article we also investigate in more detail the impact of the $Z p_T$ data, noting no particular sensitivity to the cuts used in our fit.  

We have repeated our study of a determination of the best-fit value of 
$\alpha_S(M_Z^2)$, but now both at NNLO and at aN$^3$LO \cite{Cridge:2024exf}, together with some other minor updates compared to our previous study. We now find $\alpha_S(M_Z^2)=0.1171 \pm 0.0014$ at NNLO and the completely compatible value of $\alpha_S(M_Z^2)=0.1170 \pm 0.0016$ at aN$^3$LO. The uncertainty is larger at aN$^3$LO because at this order we have an additional theoretical uncertainty which is absent at NNLO, and hence the aN$^3$LO value represents a more complete estimation of the total uncertainty with full correlation between the additional theoretical uncertainty and that normally associated with the PDFs taken into account. Moreover, we check the effect of including dijet data rather than the default of inclusive jet data, noting that at NNLO the former shifts the central value to $\alpha_S(M_Z^2)=0.1181$, but at aN$^3$LO it remains at  $\alpha_S(M_Z^2)=0.1170$, so more compatibility between approaches
is seen at aN$^3$LO. 

We have repeated the inclusion of QED corrections, including a photon PDF, into the aN$^3$LO study \cite{Cridge:2023ryv} (and into a LO study). At NLO and even at NNLO we have noted that inclusion of QED corrections leads to a significant deterioration in fit quality, e.g. $\Delta \chi^2 = 17$ at NNLO, but find that 
this becomes an essentially insignificant $\Delta \chi^2 = 3.6$ at aN$^3$LO. 
In other respects the QED corrections at aN$^3$LO produce similar effects to lower orders. The photon PDF is slightly larger, but this mirrors the slightly larger charge weighted quark distribution which generates the photon distribution. Similarly to lower orders, the QED-corrected gluon (and strange quark)
PDF is smaller due to loss of momentum to the photon, as seen in Fig.~\ref{fig:n3logluon} (left), and the high-$x$ valence quark is a little lower due to photon radiation. The slight decrease in the gluon has an effect on predicted cross-sections, e.g. gluon-gluon fusion production of a Higgs boson, as seen in Fig.~\ref{fig:n3logluon} (right). 
Additional studies related to the MSHT20 PDFs can be found in  \cite{Jing:2023isu,Cridge:2023ztj,Armesto:2023hnw}. 

\begin{figure}
\begin{center}
\includegraphics[scale=0.99]{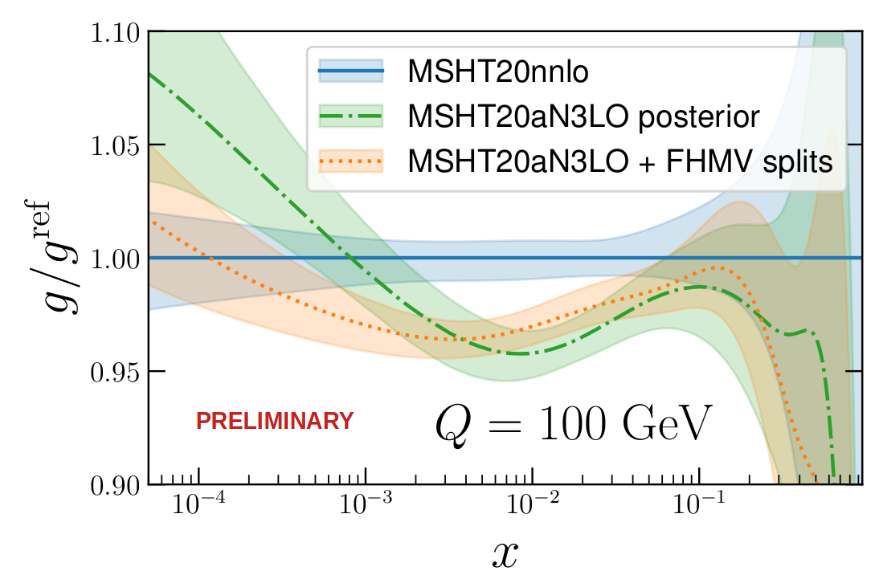}
\hspace{0.2cm}
\includegraphics[scale=0.623]{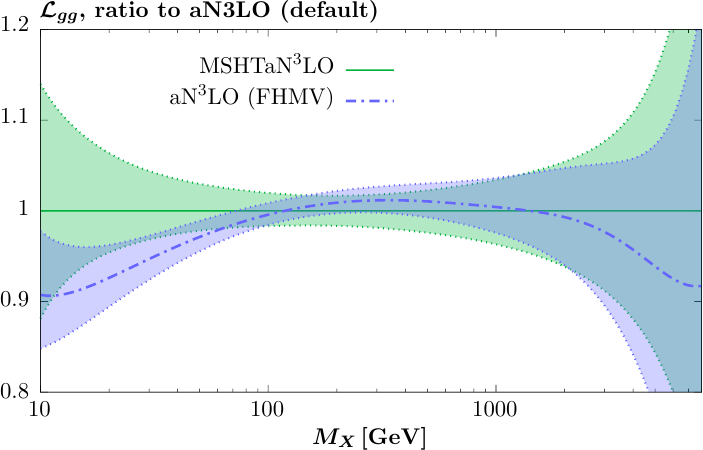}
\caption{\sf The gluon distribtion at $Q^2=10^4$~GeV$^2$ resulting from the MSHT global fit at aN$^3$LO with updated splitting functions compared to that from the published MSHT aN$^3$LO PDFs and to the MSHT NNLO PDFs (right). The gluon-gluon luminosity as a function of final state invariant mass $M_X$ for the MSHT global fit at aN$^3$LO with updated splitting functions compared to that from the published MSHT aN$^3$LO PDFs (right).}
\label{fig:newsplit}
\end{center}
\end{figure}

\section{Improvements to the aN$^3$LO PDFs}

\begin{figure}
\begin{center}
\includegraphics[scale=0.62]{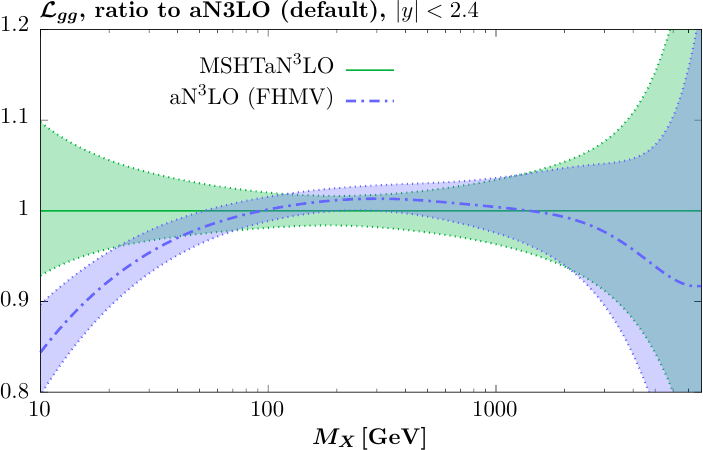}
\hspace{0.2cm}
\includegraphics[scale=0.62]{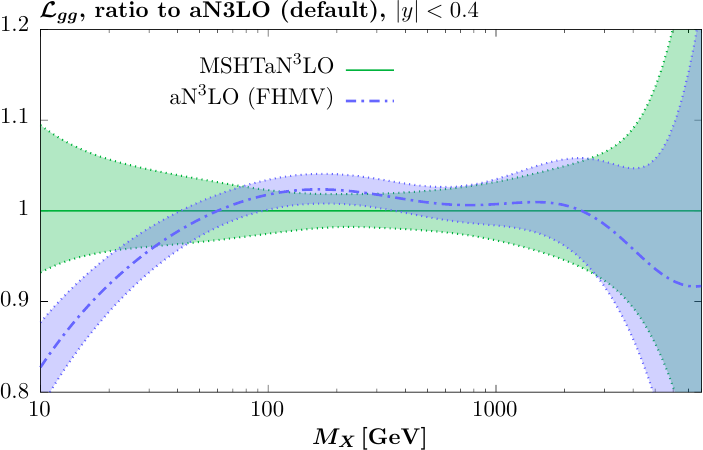}
\caption{\sf The gluon-gluon luminosity as a function of final state invariant mass $M_X$ for the MSHT global fit at aN$^3$LO with updated splitting functions compared to that from the published MSHT aN$^3$LO PDFs with a rapidity cut of $|y| < 2.4$ applied (left). The same but with $|y| < 0.4$ applied (right).}
\label{fig:lumi}
\end{center}
\end{figure}

Returning to the standard aN$^3$LO study, as mentioned earlier we have furthermore investigated the impact of including more up to date information on splitting functions, i.e. that contained in \cite{Falcioni:2023luc,Falcioni:2023vqq,Moch:2023tdj}, which makes us more directly comparable to the results of those in \cite{NNPDF:2024nan}. The result for the gluon of a refit with the central splitting functions of \cite{Moch:2023tdj} are shown in Fig.~\ref{fig:newsplit} (left), though we note that the uncertainties shown 
do not include a component due to the remaining uncertainties on the splitting functions. The use of  the central value of the improved aN3LO splitting functions changes the aN$^3$LO gluon a little compared to the published 
aN$^3$LO MSHT PDFs, raising it by about $1.5\%$ near $x = 0.01$. In other regards the main features of the comparison of the aN$^3$LO PDFs to those at NNLO remain very similar. The $\chi^2$ for the aN$^3$LO global fit is about 50 
higher than that of \cite{McGowan:2022nag}  (i.e. still more than 100 units lower than at NNLO), very largely due to differences in detail at the 
smallest $x$ values, and this would certainly improve to some extent once the uncertainty in the splitting functions is accounted for. The impact on the gluon-gluon luminosity for a centre-of-mass energy of 14~TeV is shown in Fig.~\ref{fig:newsplit} (right), and one sees that although there is some change in the luminosity compared to the published aN$^3$LO PDFs, this is 
washed out to some extent by the integration over $x$ and the increase for $M_X=125$~GeV is very small indeed. In Fig.~\ref{fig:lumi} we show the luminosity for rapidity cuts of 2.4 (left) and 0.4 (right), and we see that as the cut decreases and the range of $x$ sampled diminishes the change in the luminosity at specific values of $M_X$, particularly $M_X=125$~GeV,
becomes more distinct. Hence, it is likely that the most significant effect of improved splitting functions on predictions for the LHC will show up in the details of distributions rather than in inclusive cross sections.  A complete update of the aN$^3$LO PDF determination is an ongoing study and results will appear in a future publication. 

\end{document}